# $T_c$-maximum in Solid Solution of Pyrite IrSe$_2$-RhSe$_2$ Induced by Destabilization of Anion Dimers

Jiangang Guo,[†] Yanpeng Qi,[†] Satoru Matsuishi,[‡] Hideo Hosono[†,‡,*]

[†]Frontier Research Center, and [‡]Materials and Structures Laboratory, Tokyo Institute of Technology, Yokohama 226-8503, Japan

*Supporting Information Placeholder*

**ABSTRACT:** We have established a well-defined dome shape $T_c$ curve in Ir$_{0.94-x}$Rh$_x$Se$_2$ superconductors. The maximum $T_c^{onset}$ of 9.6K was obtained at x=0.36, at which the *Se-Se* separation in the dimer anion is the longest. Simultaneously, the destabilization of *Se-Se* dimers accompanied by partial electron-transfer from the Ir/Rh to the chalcogenide ions resulted in the emergence of optimal $T_c$.

Exploring a novel parent compound and then regulating its superconductivity has been a longstanding challenge in the materials community. The superconducting critical temperature ($T_c$) of cuprates and iron-based superconductors, although not yet clearly understood, should have intimate correlation with local structure distortion. Experimentally, their $T_c^{max}$ values are proportional to the number of CuO$_2$ layers and regular degree of FeX$_4$ (X=As, Se) tetrahedrons, respectively.[1,2] In particular, the isovalent substitution of Te/P for Se/As in iron-based superconductors demonstrated that symmetry of local atomic structure could not only drastically enhance the scale of $T_c$ as well, but could profoundly influence the normal-state transport properties.[3,4] Recently, as a structural parameter, the dimerization has been reported in the low-temperature insulting phase of three-dimensional CuIr$_2$S$_4$, which resulted in spin singlets and was responsible for the strong loss of magnetic moment.[5] Also, breaking the molecule-like anion dimer in SrCo$_2$(Ge$_{1-x}$P$_x$)$_2$ successfully induced the unexpected ferromagnetic quantum critical point between both paramagnetic end members.[6]

Pyrite TM*Ch*$_2$ (TM=transition metal; *Ch*=chalcogenide atom) is another kind of dimer-containing compound, in which strong *p-p* hybridization within the *Ch-Ch* pairs acts as divalent *Ch*$_2^{2-}$, and transition metal formally adopts the TM$^{2+}$ cation.[7,8] Its unique *Ch*$_2^{2-}$ dimer and metal-cation sublattices arrange themselves in an interpenetrating face-centered-cubic geometry (see Fig 1d). The geometry of the dimer not only strongly modifies the crystallographic structure, but also tunes the number of electrons in the high-energy orbital. Very recently, we have discovered that pyrite Ir$_x$Se$_2$ is a superconductor, of which the normal-state resistivity maintains a non-metal behavior and $T_c$ monotonously increases up to ~6 K as the scale of the unit cell and bond-length of the *Ch*$_2^{2-}$ dimer expands.[9] This discovery offers a rare opportunity to investigate the superconductivity and normal-state properties by tuning the bonding state of Se-Se dimers in pyrite compounds. In this communication, we describe a well-defined dome shape of $T_c$ responding to negative chemical pressure in Ir$_{0.94-x}$Rh$_x$Se$_2$ superconductors. The normal-state resistivity evolves from non-metal to strange metal, and then to classic Fermi liquid metal. Moreover, the specific heat capacity measurement demonstrated that Ir$_{0.58}$Rh$_{0.36}$Se$_2$ exhibits maximal $T_c$, ~9.6K, accompanied by the weakening of the dimer bonding and possible structural instability.

The starting materials—Ir, Rh and Se powders—were mixed well and loaded into the h-BN capsule in an Ar-filled glove box (O$_2$, H$_2$O<1ppm). The assembly was then subjected to a belt-type high-pressure machine with ~5GPa and ~1773K for 2h. The as-prepared samples were characterized by powder x-ray diffraction (PXRD) using a Bruker D8 Advance with Cu $K_a$ radiation. Rietveld refinements of the data were performed with the TOPAS package.[10] The electrical resistivity was measured through the standard four-wire method on the physical property measurement system (PPMS, Quantum Design). Specific heat measurement was performed by the thermal relaxation method in the temperature range of 1.8–20 K (PPMS, Quantum Design). The dc magnetic properties were characterized using a vibrating sample magnetometer (VSM, Quantum Design). The chemical compositions of the samples were examined by electron probe microscope analysis (EPMA) with backscattered electron (BSE) mode.

Figure 1a shows the measured PXRD pattern and Rietveld refinement results for Ir$_{0.58}$Rh$_{0.36}$Se$_2$ as a representative composition. Except for slight Ir metal impurity, all the reflections could be indexed with a cubic cell of parameters a=b=c=6.00365(2) Å, V=216.394(4) Å$^3$. Examination of diffraction extinction revealed that all of the samples crystallized in a primitive lattice with a probable space group of Pa$\bar{3}$ (No. 205), and the Rietveld refinement confirmed that the crystal structure was isostructural to our previous data.[9] As shown in Figure 1b, the size of the unit cell could continuously increase with the substitution of Rh for Ir. For x< 0.65, the lattice constant monotonously increased to 6.00825(2) Å and then it slowly became saturated at the higher Rh content. Following simple ionic model, their unit cell dimension should decrease with increasing Rh. Thus, this unexpected increase suggests that the effect of the crystal electric field and $Se_2^{2-}$ polarization needs to be taken into account, as pro-posed by Brikholz.[11,12] Regarding to the critical parameters, the distance of Se-Se dimers and the octahedral of $MSe_6$ (M=Ir or Rh) unit are shown in Fig 1c and Fig 1e, respectively. The bond length of the $Ch_2^{2-}$ dimer slowly increased to the maximum of ~2.65 Å and then drastically decreased for x>0.5, implying that rhodium substitution could effectively control the non-bonded and bonded state of the dimers. At the same time, we could obtain the bonding information on the $MSe_6$ octahedron from the Rietveld refinement (see Fig. S1). The Rh substitution elongated all bonds, i.e. M-Se, $Se_1$-Se and $Se_2$-Se, and substantially enlarged the size of the $MSe_6$ octahedron, which is the possible origin of anomalous lattice expansion. The DFT calculations show the width of energy band near the Fermi level gradually decreases with x (see Fig. S2). Thus, the structural analysis indicated that the transition in $Ir_{0.94-x}Rh_xSe_2$ is continuous in nature.

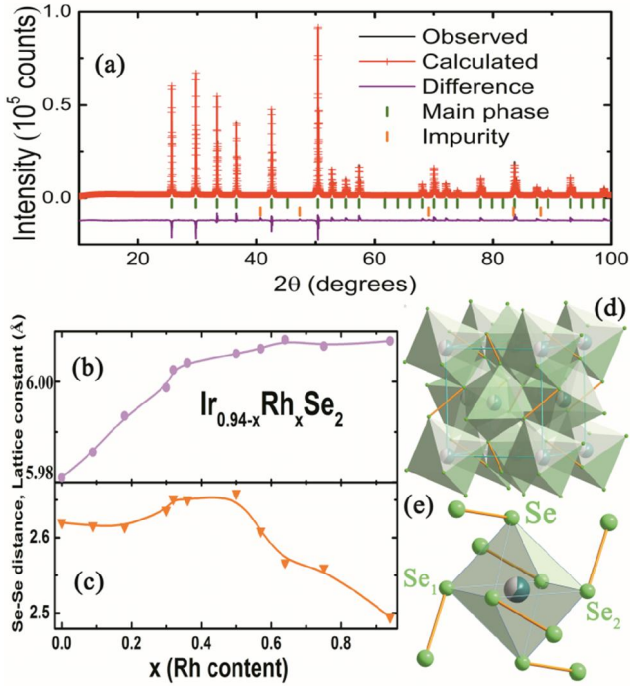

**Figure 1.** Powder x-ray diffraction pattern and crystal structure of $Ir_{0.94-x}Rh_xSe_2$. (a) Observed and fitted XRD pattern of $Ir_{0.58}Rh_{0.36}Se_2$. The result of the Rietveld refinement showed good convergence ($R_{wp}$ = 5.92%, $R_p$ = 4.53%). (b) (c) The variation of the lattice constant and the Se-Se dimer distance. (d) The crystal structure of $Ir_{0.94-x}Rh_xSe_2$ solid-state solution. The $MSe_6$ octahedral are connected by the Se-Se dimers (yellow bonds). (e) A fragment of the $MSe_6$ octahedral, where all corners are bonded with Se-Se dimers. Se-$Se_1$ and Se-$Se_2$ denote two types of Se-Se bond in one octahedron.

Figure 2a and 2b show the temperature dependence of electrical resistivity under a zero field below 305K and 12K, respectively. Let us first focus on the normal-state transport properties of $Ir_{0.94-x}Rh_xSe_2$ with 0 ≤ x ≤ 0.94. As shown in Figure 2a, the parent compound showed the negative temperature coefficient at higher temperature, suggesting its normal-state property as a non-metal or semiconductor. With further substitution, thermal coefficients of the normal-state resistivity slowly became positive, corresponding to metal behavior. We used the power-law equation $\rho(T)=\rho_0+AT^\alpha$ ($\rho_0$: residual resistivity, A: constant) to fit all normal-state resistivity curves up to 200K. The plot of the obtained value α versus rhodium composition is shown in Figure 2c. With increasing x, α gradually increases to 1 at the intermediate range, and then approaches 2 for the $Rh_{0.94}Se_2$ end member. For x=0.36, one can observe nearly perfect T-linear resistivity dependence at a wide temperature range above $T_c$, which strikingly deviates from the classic Fermi-liquid theory. It is evident that the electron-phonon scattering dominates the process of conductivity. Thus, as the increase in x, the normal-state transport property evolves from non-metal to T-linear resistivity, and finally to a classic metal.

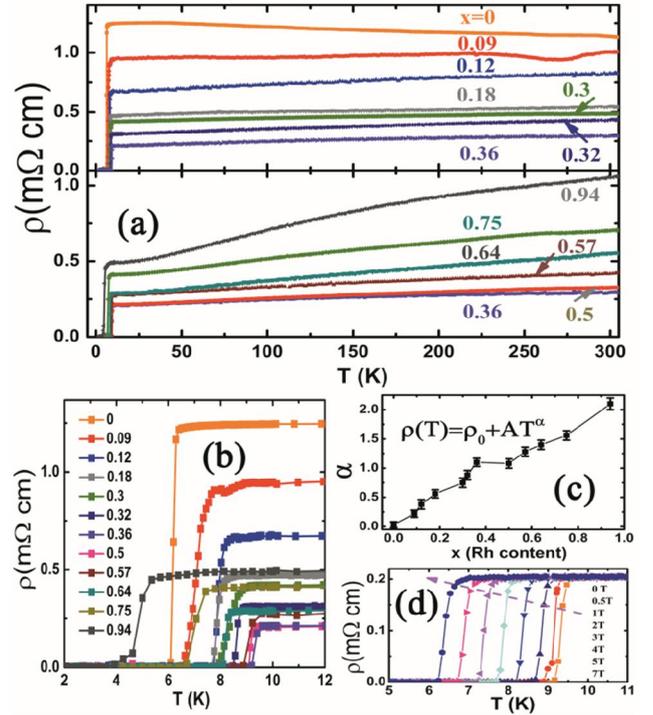

**Figure 2.** (a), (b) The temperature dependence of electrical resistivity under zero magnetic field below 305K and 12 K, respectively. (c) The value of α obtained from the fitting results of normal-state electrical resistivity. (d) The resistivity of $Ir_{0.58}Rh_{0.36}Se_2$ under various magnetic fields.

Figure 2b shows the superconducting properties for the lower-temperature zone. It can be seen that the resistivity abruptly dropped and zero resistivity was obtained. In addition, the fact that $T_c$ first increased and then decreased with increasing x is clearly shown. The maximal $T_c^{onset}$, 9.6K, occurred at x=0.36, which corresponds to the linear-resistivity temperature dependence. The $T_c$ of $RhSe_2$ agrees with the previous report.[13] It is worth noting that the anomalous T-linear resistivity behavior is seen for the region of the maximal $T_c$. This correspondence between the region at which T-linear dependence of ρ in the normal state and the optimal $T_c$ region has been reported in cuprates and Iron-based superconductors.[14,15] Figure 2d shows the resistivity of $Ir_{0.58}Rh_{0.36}Se_2$ under different magnetic fields. The transition was suppressed by applying external magnetic fields. The estimated upper critical fields $H_{c2}$ were 28.5T and 24.5T from the WHH model[16] and Ginzburg-Landau equation, respectively (see Fig. S3c).

The enhanced superconductivity was further demonstrated by the low-temperature magnetic susceptibility and heat capacity. Figure 3a and 3b show that a large diamagnetic signal—evidence of bulk superconductivity—was observed in all samples. The variation and the maximal value of $T_c$ were consistent with the above electrical measurement. The estimated lower critical field $H_{c1}$ was about 200Oe from the isothermal magnetization curve, indicating that the material is a type-II superconductor (See Fig. S3).

Further investigation of the superconducting phase and lattice vibration were carried out by heat capacity measurement. From Figure 3c, we can see that the large superconducting jump for x=0.36 coincides with the $\rho(T)$ and $\chi(T)$ data. We could easily fit these curves above $T_c$ with the Debye model $C_p/T=\gamma+\beta T^2$, in which $\gamma$ is the Sommerfeld electronic coefficient and the other term is the lattice contribution. From the intercept of plots, we found that the electronic coefficients are 6.41 mJ·mol$^{-1}$·K$^{-2}$, 8.02 mJ·mol$^{-1}$·K$^{-2}$, 3.57 mJ·mol$^{-1}$·K$^{-2}$ for x=0, 0.36, 0.94, respectively. Following the free electron model, the density of states (DOS) at Fermi energy, $N(E_F)=3\gamma/\pi^2 K_B^2$, are calculated as 2.64 states·eV$^{-1}$·f.u.$^{-1}$, 3.3 states·eV$^{-1}$·f.u.$^{-1}$, and 1.47 states·eV$^{-1}$·f.u.$^{-1}$, respectively. Density functional calculations (see Fig. s2) shows the Fermi energy of $Ir_{0.5}Rh_{0.5}Se_2$ locates at the peak of DOS, which is consistent with this observed data.

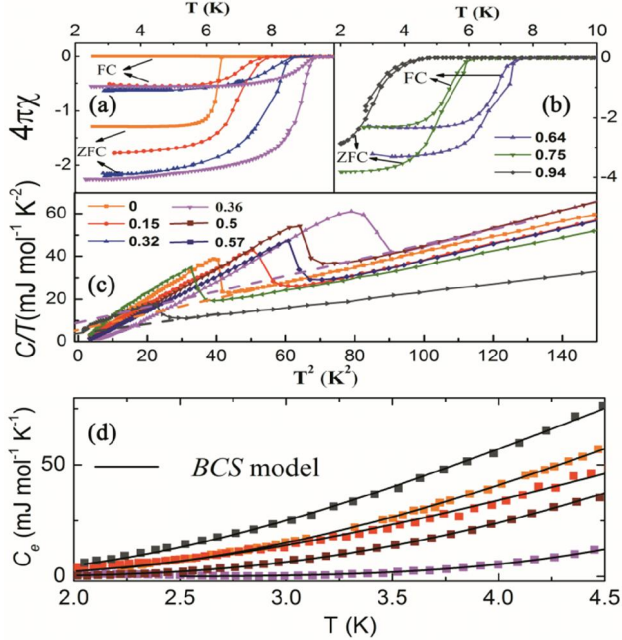

**Figure 3.** (a) and (b) Temperature dependences of magnetic properties at a magnetic field H of 10Oe for $Ir_{0.94-x}Rh_xSe_2$ at zero-field-cooling (ZFC) and field-cooling (FC) process. (c) The total heat capacity divided by T as a function of temperature squared. The dashed lines are the extrapolated line from Debye model fitting. (d) Temperature dependence of the electronic specific heat; the solid line is the conventional BCS model fitting result.

The value of slopes $\beta$ was largest value (0.41 mJ·mol$^{-1}$·K$^{-4}$) for x=0.36. We plotted $\Theta_D$, evaluated from the equation $\Theta_D=(12\pi^4 nR/5\beta)^{1/3}$, as a function of x in Figure 4b. The $\Theta_D$ of intermediate compositions reached a minimum value, ~242K. Since the Debye frequency is proportional to the Debye temperature, we deduced there may be an occurrence of phonon softening from the reduced frequency. The softer phonon mode possibly implies that the crystal structure approached the structural instability states.

In Figure 3d, we show the electronic-specific heat $C_e=C_{total}-C_{phonon}$ versus $T$, where $C_{phonon}$ comes from the Debye model results. The s-wave BCS fitting of $C_e$ vs T curves[17] gave $2\Delta/k_B T_c = 5.95$ for x=0.36, suggesting that the coupling state was the strongest for the weaken dimer. As the dimers close, the whole system gradually transits to the classic BCS weak coupling state (see Fig. S4). Moreover, the normalized heat capacity jump $\Delta C/\gamma T_c$ exhibited a similar trend, and also reached the maximal value 3.55 at x=0.36, as shown in Figure 4b. Therefore, we could safely infer the rarely critical nature by modifying the bonding state of the dimers, indicative of $T_c$, $\gamma$, and the Debye temperature in $Ir_{0.94-x}Rh_xSe_2$.

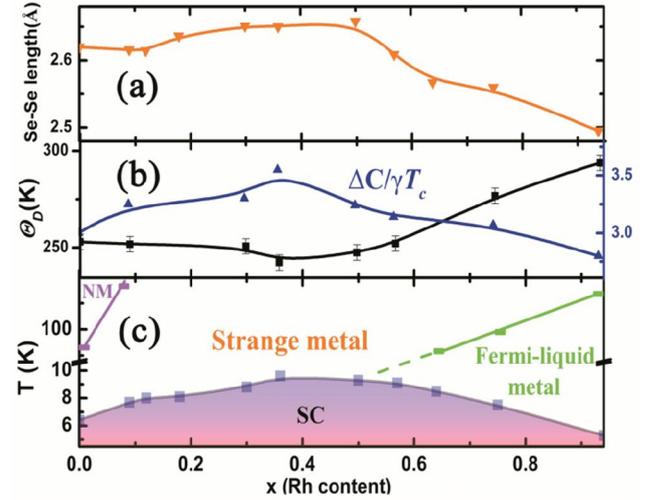

**Figure 4.** Summary and phase diagram of $Ir_{0.94-x}Rh_xSe_2$. (a) The variation of the Se-Se dimer bond length. (b) The Debye temperature $\Theta_D$ and $\Delta C/\gamma T_c$ as a function of rhodium content x. (c) The electronic phase diagram of $Ir_{0.94-x}Rh_xSe_2$. The violet symbol denotes the crossover temperature from the non-metal (NM) to the strange-metal state. The green symbol denotes the crossover temperature from the strange-metal to the classic Fermi-liquid-metal state.

The main results are summarized in Fig.4 as an electronic phase diagram. With increasing bond length of dimer, both the $\Theta_D$ and $T_c$ of $Ir_{0.94-x}Rh_xSe_2$ slowly approach their extremes. It should be noted that at the lowest Debye temperature, the largest $\Delta C/\gamma T_c$ and $T_c$ emerge at the vicinity of the weakening dimer states. As the dimer bonding is enhanced again, $T_c$ starts to decrease and the high-frequency lattice vibration is recovered. The present data clearly shows that the bonding strength in dimer states in $Ir_{0.94-x}Rh_xSe_2$ is correlates with a dome-shaped $T_c$ curve and significantly influenced the lattice vibration. It has been reported that $BaNi_2(As_{1-x}P_x)_2$ shows sudden phonon softening when the lower-$T_c$ triclinic phase was converted to a higher-$T_c$ tetragonal phase by isovalent substitution, even though the $N(E_F)$ was independent of doped content.[18] Furthermore, the $T_c$ of $BaSi_2$ could be enhanced from 6K to 8.9K by flattening the Si planes, which appeared at the structural instability state involving softening of the Si phonon mode.[19] In the present case, the combined results demonstrated that the charge fluctuation between the $Se_2^{2-}$ dimer and cation essentially rise as the dimer bonding gradually weaken, which is responsible for to the increment of $\gamma$. The PDOS pattern around Fermi energy shows that the contribution of *Se* orbital increases whereas the part of transition metal almost keeps constant with the increasing of Ir content, indicating the hybridization state of Ir/Rh-*d* and Se-4*p* orbital becomes stronger. Although the DOS shows the maximal value at the x=0.36, there could be another possibilities resulting in the significantly enhancement of $T_c$ (superconducting gap), such as strengthening of electron-phonon coupling and softening of phonon due to the structural instability at the edge of weak dimers states. Further phonon spectrum calculations will be required to validate the proposed mechanism.

To summarize, we found a domed $T_c$ curve in $Ir_{0.94-x}Rh_xSe_2$ by tuning the bonding state of $Se_2^{2-}$ dimers. At the optimal $T_c$, the *Ch-Ch* separation, electronic density of states and $\Delta C/\gamma T_c$ all reach their maximal values. In addition, the normal-state resistivity shows T-linear dependence in the range of the weakening dimer, and then approaches conventional metal with the shortening of the dimer. These results suggest that a transition metal pyrite is a good platform for superconductivity to emerge from modification of the bonding state of the dimer anions.

## ASSOCIATED CONTENT

### Supporting Information

The bond parameters of $MSe_6$ octahedra, the isothermal magnetization and critical fields of $Ir_{0.58}Rh_{0.36}Se_2$, fitting results of $C_e$ vs T curves and density functional theory calculation detail and results. This material is available free of charge via the Internet at http://pubs.acs.org.

## AUTHOR INFORMATION

### Corresponding Author

hosono@msl.titech.ac.jp

### Notes

The authors declare no competing financial interests.

## ACKNOWLEDGMENT

We acknowledge Dr. H. Lei and Mr. S. Iimura for the valuable discussion. This work was supported by the Funding Program for World-Leading Innovative R&D on Science and Technology (FIRST), Japan.

# $T_c$-maximum in Solid Solution of Pyrite IrSe$_2$-RhSe$_2$ Induced by Destabilization of Anion Dimers

Jiangang Guo,[†] Yanpeng Qi,[†] Satoru Matsuishi,[‡] Hideo Hosono[†,‡,*]

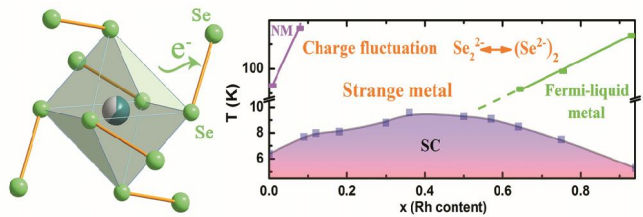